%Paper: hep-lat/9210004
%From: Kondo <kondo@tansei.cc.u-tokyo.ac.jp>
%Date: Fri, 2 Oct 92 11:47:44 JST
%Date (revised): Tue, 13 Oct 92 00:31:13 JST

%This needs macrosharvmac
% and EPmacros.tex(included).
%
%Three postscript figures are included
%as fig1.ps, fig2a.ps and fig2b.ps.
%Please split the file from the end of this file.
%
%%%%%%%%%%%%%%%%%%%%%%%%%%%%%%%%%%%%%%%
\input harvmac
%\input epsf

%\input EPmacros
%\input EP61-macab

%%%%%%%%%%%%%%%%%%%%%%%%%%%%%%%%%%%%%%%%%%%%%%%%%%%%%%%%%%%%%%%%%%%%%%%%

% ***************** Macro file [EPmacros.tex] *********************
%---------------------------------------------------------------------
%
  %     curly letters

\def\vev#1{\langle #1 \rangle}

\def\darr#1{\raise1.5ex\hbox{$\leftrightarrow$}\mkern-16.5mu #1}
 %pound sterling
\def\ha{{1\over2}}
\def\half{{\textstyle{1\over2}}} %puts a small half in a displayed eqn
\def\roughly#1{\raise.3ex\hbox{$#1$\kern-.75em\lower1ex\hbox{$\sim$}}}
%%%%%--------------------------------------------------
\def\today{\ifcase\month\or
  January\or Febrary\or March\or April\or May\or June\or
  July\or August\or September\or October\or November\or December\fi
  \space\number\day, \number\year}
%%%%%--------------------------------------------------
\def\gtnoteq{\mathrel{\hbox{\raise0.2ex\hbox{$>$}
  \kern-0.75em\raise-0.9ex\hbox{$\not=$}}}}

%\def\bpsi{\bar\psi}
%\def\intDe#1{ \int_{\De{#1}} }

%%%%%--------------------------------------------------
% The following is for Mac
%\def\graph{\lower2cm
% \vbox to 4.5cm{\hrule width 4.8cm height 0pt depth 0pt
%       \vfill\special{illustration pic.ps}}}
%%%%%--------------------------------------------------

%%%% Some definition from the PHYZZXY.TeX.
%%%%%%%%%%%%%%%%%%%%%%%%%%%%%%%%%%%%%%%%%%%%%%%%%%%%%%%%%%%%%%%%%%%%%%%%
%   Miscellaneous macros
%

%
     
\def\eg{\hbox{\it e.g.}}     

\def\\{\relax\ifmmode\backslash\else$\backslash$\fi}
\def\globaleqnumbers{\relax\if\equanumber<0\else\global\equanumber=-1\fi}

\def\journal#1&#2(#3){\unskip, \sl #1~\bf #2 \rm (19#3) }

\def\Tr{\mathop{\rm Tr}\nolimits}
\let\int=\intop		
\def\prop{\mathrel{{\mathchoice{\pr@p\scriptstyle}{\pr@p\scriptstyle}{
		\pr@p\scriptscriptstyle}{\pr@p\scriptscriptstyle} }}}
\def\pr@p#1{\setbox0=\hbox{$\cal #1 \char'103$}
   \hbox{$\cal #1 \char'117$\kern-.4\wd0\box0}}
\def\lsim{\mathrel{\mathpalette\@versim<}}
\def\gsim{\mathrel{\mathpalette\@versim>}}
\def\@versim#1#2{\lower0.2ex\vbox{\baselineskip\z@skip\lineskip\z@skip
  \lineskiplimit\z@\ialign{$\m@th#1\hfil##\hfil$\crcr#2\crcr\sim\crcr}}}
%
%%%%%%%%%%%%%%%%

%%%%%%%%%%%%%%%%%%%%%%%%%%%%%%%%%%%%%%%%%%%%%%%%%%%%%%%
%%                                                   %%
%%   CHIBA-EP-XX.TEX general definitions             %%
%%                                                   %%
%%%%%%%%%%%%%%%%%%%%%%%%%%%%%%%%%%%%%%%%%%%%%%%%%%%%%%%

%%%%%picture definition
\def\picture #1 by #2 (#3){
  \vbox to #2{
    \hrule width #1 height 0pt depth 0pt
    \vfill
    \special{picture #3}}}

\def\scaledpicture #1 by #2 (#3 scaled #4){{
  \dimen0=#1 \dimen1=#2
  \divide\dimen0 by 1000 \multiply\dimen0 by #4
  \divide\dimen1 by 1000 \multiply\dimen1 by #4
  \picture \dimen0 by \dimen1 (#3 scaled #4)}}
%%%%%picture definition end

%%Journals[References Macros]

\def\NPB#1{Nucl.\ Phys. {\bf B{#1}}}

\def\PLB#1{Phys.\ Lett. {\bf B{#1}}}

\def\PRL#1{Phys.\ Rev.\ Lett. {\bf {#1}}}
\def\PTP#1{Prog.\ Theor.\ Phys. {\bf {#1}}}

%%

%%Greek Letters %
\def\Ga{\alpha}        %\def\GA{\alpha}
\def\Gb{\beta}         %\def\GB{\beta}
\def\Gd{\delta}        
      
\def\Gf{\phi}          \def\GF{\Phi}
\def\Gg{\gamma}        

\def\Gk{\kappa}        %\def\GK{\kappa}
\def\Gl{\lambda}       
\def\Gm{\mu}           %\def\GM{\mu}
\def\Gn{\nu}           %\def\GN{\nu}
           \def\GP{\Pi}
        
\def\Gr{\rho}          %\def\GR{\rho}
\def\Gs{\sigma}        

        \def\GW{\Omega}
           
\def\Gy{\psi}          
         %\def\GZ{\zeta}
%%

%%Character square bracket [ and ].%

\def\MKO{\Bigl[}
\def\MKC{\Bigr]}

%%Character brace { and }.  %

\def\MCO{\Bigl\{}
\def\MCC{\Bigr\}}

%%Character bracket ( and ).%

\def\MO{\Bigl(}
\def\MC{\Bigr)}

\def\AO{\langle}
\def\AC{\rangle}

%%
%%%%%%%%%%%%%%%%%%%%%%%%%%%%%%%%%%%%%%%%%%%%%%%%%%%%%%%
%%                                                   %%
%%          put your own definitions here            %%
%%                                                   %%
%%%%%%%%%%%%%%%%%%%%%%%%%%%%%%%%%%%%%%%%%%%%%%%%%%%%%%%

%%Math. Functions %
\def\Tr{\mathop{\rm Tr}\nolimits}
\def\tr{\mathop{\rm tr}\nolimits}
\def\Det{\mathop{\rm Det}\nolimits}

\def\to{\rightarrow}

\def\pd{\partial}

\def\para#1{\par
   \ifnum\the\lastpenalty=30000\else \penalty-100\smallskip \fi
   \noindent{#1}\enspace \vadjust{\penalty5000}}

\def\oneh{1}
\def\twoh{2}
\def\ha{{1\over2}}
\def\half{{\textstyle{1\over2}}} %puts a small half in a displayed eqn
\def\bone{{\bf 1}}

\def\Gyb{{\bar\psi}}
\def\GFd{\GF^{\dag}}
%%

%%Special Math. Function %
%
\def\BH{\kappa}
\def\BY{y}
\def\BW{w}

\def\vev#1{\langle #1 \rangle}
\def\Hvev#1{{\langle #1 \rangle}_H }
\def\HVEV#1{{\left\langle #1 \right\rangle}_H }
\def\GFd{\GF^{\dag}}

\def\GFh{{\hat \GF}}
\def\GFdh{{\hat \GF^{\dag}}}

\def\HVEVtrFFh#1#2{{\left\langle\tr\GFdh_{ #1}\GFh_{ #2}\right\rangle}_H }
\def\HVEVtrFFFFh#1#2#3#4{{\left\langle
          \tr\GFdh_{ #1}\GFh_{ #2}\GFdh_{ #3}\GFh_{ #4}\right\rangle}_H }
%%special modification for CHIBA-EP-61 and -62 by T.E.[050992]
%%        tr -> tr_G
%\def\trG{\tr_G}
%\def\VEVtrFF#1#2{\left\langle\trG\GFd_{ #1}\GF_{ #2}\right\rangle}
%\def\VEVtrFFFF#1#2#3#4{\left\langle
%          \trG\GFd_{ #1}\GF_{ #2}\GFd_{ #3}\GF_{ #4}\right\rangle}
%\def\HVEVtrFF#1#2{{\left\langle\trG\GFd_{ #1}\GF_{ #2}\right\rangle}_H }
%\def\HVEVtrFFFF#1#2#3#4{{\left\langle
%          \trG\GFd_{ #1}\GF_{ #2}\GFd_{ #3}\GF_{ #4}\right\rangle}_H }

%%

%%Special Letter %
\def\QeD{Q\kern-.2em\lower.5ex\hbox{E}\kern-.06em D}
\def\QcD{Q\kern-.2em\lower.5ex\hbox{C}\kern-.06em D}

\def\eff{{\rm eff}}

%

%%%%%%%%%%%%%% Definition end %%%%%%%%%%%%%%%%%%%%%%%
\message{ Go ahead, please.(by ET)}

%*****************   The end of Macro file    *********************

%%%%%%%%%%%%%%% TITLE PAGE %%%%%%%%%%%%

\Title{\vbox{\baselineskip12pt
 \hbox{CHIBA--EP--61-REV}
% \hbox{\today}
 }}
{{\vbox{
\centerline{
 Multicritical Points in a Lattice Yukawa Model
 }\vskip10pt
\centerline{
 with the Wilson-Yukawa Coupling $^\ast$
 } }}}
\centerline{
 Toru Ebihara  $^\dagger$
 and Kei-ichi Kondo $^\ddagger$
 }
\bigskip
\centerline{${}^\dagger$
 Graduate School of Science and Technology, ~Chiba University, }
\centerline{ 1-33 ~Yayoi-cho,  ~Inage-ku, ~Chiba 263, JAPAN }
\smallskip
\centerline{${}^\ddagger$
 Department of Physics, Faculty of Science, ~Chiba University, }
\centerline{ 1-33 ~Yayoi-cho, ~Inage-ku, ~Chiba 263, JAPAN  }

\vskip1cm

  For a lattice regularized chiral-invariant $SU(2)_L\times~SU(2)_R$
fermion-scalar model with a Yukawa coupling $\BY$ and a Wilson-Yukawa
coupling $\BW$,  we investigate the phase structure and in particular
show the existence of the multicritical line, in the strong Yukawa
and/or Wilson Yukawa coupling region, at which four phases meet.
 The result is in good agreement with the Monte Carlo simulation.
 This analytical result is derived from the effective scalar model
obtained by integrating out the fermion field where the action is
explicitly obtained from the hopping parameter expansion up to
next-to-leading order.
 For estimates on the correlation function of the scalar field
we apply the mean-field method.

\vskip 1cm
\noindent

\footnote{}{$\dag$  E-mail address:
  ebihara@cuphd.chiba-u.ac.jp; f36598@jpnkudpc.bitnet}
\footnote{}{$\ddag$ E-mail address:
  c32295@jpnkudpc.bitnet}
\footnote{}{$\ast$ To be published in Physics Letters B.}

%\draft
%\Date{5/92} %Revised at
\Date{7/92}

%%%%%%%%%%%%%%%%%  Here comes the text .....

%\newsec{Introduction}

  In the standard model of electroweak interactions
the Yukawa couplings of quarks and leptons to the Higgs scalar field
provide the masses of the elementary fermions.
In particular, these masses are generated through the spontaneous
symmetry breaking.
However the formulation of the Yukawa models on the lattice suffers from
the fermion doubling problem %:
 [\ref\NN{ H.B. Nielsen and M. Ninomiya,
  \NPB{185} (1981) 20, ibid. {\bf B185} (1981) 173.}]:
the naively regularized Dirac fermion on the lattice describes additional
fermion degrees of freedom (doubler fermions) besides one fermion desired
in the continuum limit.
To avoids this problem in the chiral theories one must remove unwanted
doublers in the continuum limit in such a way that the chiral symmetry is
preserved on the lattice %.
 [\ref\AFK{ S. Aoki, \PRL{60} (1988) 2109;
 K. Funakubo and T. Kashiwa, ibid. {\bf 60} (1988) 2113.},
 \ref\SS{J. Smit, Acta Physica Polonica {\bf B17} (1986) 531;
 P.V.D. Swift, \PLB{145} (1984) 256.}].
One formulation of the Yukawa model adds a manifestly chiral invariant
Wilson-Yukawa coupling to decouple the doublers from the physical
spectrum by rendering them heavy dynamically [\SS].
The other method introduces mirror fermions explicitly, in addition
to the original field, which transform oppositely under
the chiral symmetry group %.
 [\ref\Montvay{ I. Montvay, DESY preprint, DESY-92-001.}].

  In this letter we examine the former one, lattice Yukawa model
with a chiral invariant Wilson-Yukawa term with coupling constant $\BW$,
which reduces to the naive lattice Yukawa model with only a Yukawa
coupling $\BY$ in the limit  $\BW\to0$.
The naive lattice Yukawa model ($\BW=0$) which is defined below has a
rich phase structure as shown in %Fig.
 \fig\PDnoW{
 Phase diagram of the naive Yukawa model ($\BW=0$)
 for $N_f=4$, $d=4$. },
which is already
discussed in the previous paper %.
 [\ref\EK{T. Ebihara and K.-I. Kondo,
 \PTP{89} (1992) 1019.}].
To distinguish the phase, two order paramters have been  introduced:
one is the usual expectation value of the scalar field $\Gf_x$ and another
is the expectation value of the staggered scalar field defined by
$\hat\Gf_x :=~(-1)^{x_1+x_2+\cdots+x_d} \Gf_x$ where $d$ denotes the
dimension of the Euclidean space.
The phase diagram consists of
the ferromagnetic (FM) phase
($\vev{\Gf}\neq~0$, $\vev{\hat\Gf}=~0$),
the paramagnetic (PM) phase
($\vev{\Gf}=~0, \vev{\hat\Gf}=~0$),
the antiferromagnetic (AFM) phase
($\vev{\Gf}=~0, \vev{\hat\Gf}\neq~0$)
and the ferrimagnetic (FI) phase
($\vev{\Gf}\neq~0, \vev{\hat\Gf}\neq~0$).
One of the most characteristic features of this phase diagram is that
PM and AFM phases are separated into their weak (W) and strong coupling
regions (S) by the presence of the intermediate FI phase.
For $\BW=0$ the various phase transition lines intersect at the two
quadruple points $A$ and $B$ where four phases can meet.

  At a nonzero value of the Wilson-Yukawa coupling $\BW\neq~0$
the phase diagram can be {\it approximately} obtained by
shifting in %Fig.
\PDnoW\ the $\BY=~0$ axis by an amount of $d\BW$
in the positive $\BY$-direction, as confirmed by various results of
numerical simulations %.
 [\ref\Bocki{ W. Bock, A.K. De, K. Jansen,
 J. Jers\'ak, T. Neuhaus  and J. Smit,
 \NPB{344} (1990) 207. }].
This is indeed understood from the leading result of the hopping
parameter expansion given below.
In the 3-dimensional bare parameter space $(\BY,\BW,\BH)$
the two points $A$ and $B$ become two lines which we shall again label
by $A$ and $B$ respectively.

  Quite recently Bock et al.
 [\ref\Bockii{ W. Bock, A.K. De, C. Frick, K. Jansen
 and T. Trappenberg,
 \NPB{371} (1992) 683.}]
have shown that in the weak coupling region
including the point $A$ the unwanted fermion doublers can not
be removed from the physical particle spectrum.
This implies that in the weak $\BW$ region the continuum limit is
not appropriate for describing the electroweak standard model.
At least for $w=0$, moreover,
the existence of a nontrivial fixed point at A is quite unlikely from the
result of the simulation for the renormalized Yukawa coupling
[\ref\BDFJT{W. Bock, A.K. De, C. Frick,  J. Jers\'ak and T. Trappenberg,
HLRZ-91-82/ITFA-91-33;
C. Frick, talk given at the Conf. Lattice'91, Tsukuba, Japan.}] .
Thus the weak coupling region in the neighborhood of the origin
$\BY=~\BW=~0$  is governed by the Gaussian fixed point up to the point
$A$.
For $\BW=0$, J\"ulich group [\Bockii]
has also shown that the fermions become massive due to the strong Yukawa
coupling $\BY$ near the point $B$, which prevents the theory from having
the meaningful continuum limit.
Taking into account these facts, the physically meaningful point in the
continuum limit is expected to reside in the vicinity of the
multiple (would-be multicritical) point $B$  in the weak
naive-Yukawa coupling $\BY\ll~1$ and strong Wilson-Yukawa coupling
$\BW\sim~1$.

  This paper is the first to show by an analytical method that even in
the presence of the Wilson-Yukawa coupling there are multicritical
points, corresponding to $B$ in the naive case $\BW=0$,  in the strong
Yukawa and/or strong Wilson-Yukawa coupling region.
This might corresponds to a nontrivial fixed point at which an
interacting continuum limit can be realized.
Moreover in the presence of the Wilson-Yukawa coupling the doubler mass
can get mass of the order of the cutoff and therefore they are removed
from the physical spectrum in the continuum limit.
Basic ingredients are the hopping parameter expansion in higher orders
and the estimates on the expectation value of the scalar fields, of which
the mean field is the simplest one.

%\newsec{The Model}

  The chiral $SU(2)_L\times~SU(2)_R$ symmetric Yukawa model on the
lattice is defined by the action
%\eqn\Action{
%S = S_H + S_F + S_Y + S_W , }
$S = S_H + S_F + S_Y + S_W$
with
\eqn\Actions{
\eqalign{
%\eqn\ActionH{
S_H &= -\BH \sum_{x,\Gm} \half\tr\MO
   \GFd_x \GF_{x+\Gm}+\GFd_x \GF_{x-\Gm} \MC , \cr %}
%\eqn\ActionF{
S_F &= \half \sum_{x,\Gm} \MCO
   \Gyb_x \Gg_{\Gm} P_L (\Gy_{x+\Gm}-\Gy_{x-\Gm})
  +\Gyb_x \Gg_{\Gm} P_R (\Gy_{x+\Gm}-\Gy_{x-\Gm}) \MCC , \cr %}
%\eqn\ActionY{
S_Y &= \BY\sum_x\Gyb_x(\GF_x P_R+\GFd_x P_L)\Gy_x   , \cr %}
%\eqn\ActionW{
%\eqalign{
S_W &= \BW\sum_{x, \Gm}  \MCO
 \Gyb_x (\GF_x P_R+\GFd_x P_L)\Gy_x   \cr %
 &-\half \Gyb_x(\GF_x P_R +\GFd_{x+\Gm} P_L )\Gy_{x+\Gm}
 -\half \Gyb_x (\GF_x P_R +\GFd_{x-\Gm} P_L )\Gy_{x-\Gm}
 \MCC , \cr } }
where $\BH$ is the hopping parameter for the scalar field, $\BY$ the
normal Yukawa coupling, $\BW$ the Wilson-Yukawa coupling, and $P_L$
and $P_R$ are the left- and right-handed chiral projectors,
$P_L=\half (1-\Gg_5)$,  $P_R=\half (1+\Gg_5)$.
Here the scalar field $\GF_x$ is radially fixed  $\GF_x \GF_x^\dagger=~1$
and is expressed by $SU(2)$ matrix and the fermion field $\Gy$ and $\bar
\Gy$ are $SU(2)$ doublets.
  The action has the global chiral $SU(2)_L\times~SU(2)_R$ symmetry:
$\Gy_x\to~(\GW_L P_L +~\GW_R P_R )~\Gy_x$,
$\Gyb_x\to\Gyb_x~(\GW^{\dag}_L P_L +~\GW^{\dag}_R P_R )$,
$\GF_x\to\GW_L\GF_x \GW^{\dagger}_R$,
where $\GW_{L,R}\in~SU(2)_{L,R}$ respectively.

  It is convenient to transfer into the newly defined fermion field:
$\Gy_x' $ $ = [\BY+d\BW]^{-1/2}~(\GF^{\dagger}_x P_L+P_R)~\Gy_x$,
$\Gyb_x'= [\BY+d\BW]^{-1/2}~\Gyb_x~(\GF_x P_R+P_L).$
Then the fermionic part of the action \Actions\ %\Action\
is rewritten as follows:
\eqn\ActionFYW{
 S_F+S_Y+S_W= \sum_x \Gyb_x' \Gy_x'
 - K \sum_{x,y}\Gyb_x' Q_{x,y}(\GF,\GF^{\dag} )\Gy_y',}
where the matrix $Q_{x,y}$ is defined by
\eqn\MatrixQ{
Q_{x,y}=\sum^d_{\Gm =1}
  [\BW-\Gg_{\Gm}(\GFd_x\GF_{x+\Gm} P_L + P_R)] \Gd_{y,x+\Gm}
+ \sum^d_{\Gm =1}
  [\BW+\Gg_{\Gm}(\GFd_x\GF_{x-\Gm} P_L + P_R)] \Gd_{y,x-\Gm} ,}
with a hopping parameter
\eqn\i{
 K = {1\over 2(\BY+d\BW)} .}
Here we notice that the fermionic part of the action has the
bilinear form in the fermion field.
%\eqn\ii{
% S_F+S_Y+S_W = \sum_{x,y}\Gyb_x' M_{x,y}(\GF,\GF^{\dag} )\Gy_y'
% ,}
After integrating out the fermion field, the
model is converted into the effective scalar model with the action:
\eqn\EffAction{
S_{\eff}=S_H - N_f\ln\Det M ,}
where $N_f$ is the number of fermion species and
the matrix $M_{x,y}$ is defined by
\eqn\MatrixM{
M_{x,y}(\GF ,\GF^{\dag})
= \MO\bone - K  Q(\GF,\GF^{\dag} )\MC_{x,y}.}

  Now we perform the hopping parameter
expansion (HPE) %.
 [\ref\ALS{ S. Aoki, I-H. Lee and R.E. Shrock,
 \NPB{355} (1991) 383.}].  % preprint ITP-SB-90-35.}]
The first step is to expand $\ln\Det~M$  in powers of the
hopping parameter $K$ as follows:
\eqn\HPE{
\ln \Det M_{x,y}  = \Tr\ln [\bone - K Q]_{x,y}
= -\sum^{\infty}_{n=1}{1\over 2n} K^{2n} \!\!\!\!
 \sum_{x_1,\ldots,x_{2n}}\!\!\!\tr[Q_{x_1,x_2}\cdots Q_{x_{2n},x_1}] ,}
where $\tr$ should be understood over the space of the Dirac spinor and
the group $SU(2)$.

  Our result of the HPE up to $O(K^4)$ is
\eqn\EffectiveActioni{
\eqalign{
&\ln\Det~M = -{1\over2}K^2 D\sum_{x,\Gm}\MO
 2\BW^2 \tr (1) -2\tr [\GF^{\dag}_x \GF_{x+\Gm}] \MC \cr %
&-{1\over4}K^4 D \sum_{x,\Gm}\MO
 2\BW^4 (7 d -4) \tr (1)
 - 12\BW^2 \tr [\GF^{\dag}_x \GF_{x+\Gm}] \MC \cr %
 &-{1\over4}K^4 D \sum_{x,\Gm}\MO
 4\tr [\GF^{\dag}_x \GF_{x+\Gm}\GF^{\dag}_x \GF_{x-\Gm}]
 +2\tr [\GF^{\dag}_x \GF_{x+\Gm}\GF^{\dag}_x \GF_{x+\Gm}] \MC \cr
&-{1\over4}K^4 D \sum_{\scriptstyle x,\Gm,\Gn
                          \atop \scriptstyle \Gm\neq\Gn}\MO
 8\tr [\GF^{\dag}_x \GF_{x+\Gm}\GF^{\dag}_{x+\Gm+\Gn}\GF_{x+\Gm}]
 -4\tr [\GF^{\dag}_x \GF_{x+\Gm}\GF^{\dag}_{x+\Gm+\Gn}\GF_{x+\Gn}]
 \MC ,\cr }}
where $D=2^{d/2}$ for even $d$.
Here note that the trace is taken over the group $G$ and that this
result is independent of the choice of $G$.
 In the next step we specify the group: $G=SU(2)$.
The radially fixed scalar field $\GF_x$ is a $2\times~2$
$SU(2)$ matrix and therefore can be written as
$\GF=T_{\Ga}\Gf^{\Ga} \,\,  (\Ga= ~0,1,2,3)$
and its conjugate as $\GF^{\dagger}={\bar T_\Ga}\Gf^\Ga$  where
\eqn\Operator{
T_\Ga = (\Gs_0, i \Gs_m), \ \bar T_\Ga = (\Gs_0, -i \Gs_m), \
(\Ga=0,1,2,3; \ m=1,2,3)}
with a unit $2\times~2$ matrix $\Gs_0$ and Pauli matrices $\Gs_m$.
By making use of the following identities for traces:
\eqn\Identityi{
\half\tr[T^{\Ga}{\bar T}^{\Gb}]=\Gd^{\Ga,\Gb},}
\eqn\Identityi{
\half\tr[T^{\Ga} {\bar T}^{\Gb} T^{\Gr} {\bar T}^{\Gs}]
=\Gd^{\Ga,\Gb} \Gd^{\Gr,\Gs} - \Gd^{\Ga,\Gr} \Gd^{\Gb,\Gs}
+\Gd^{\Ga,\Gs} \Gd^{\Gb,\Gr} ,}
final expression of $\ln\Det~M$ is obtained as follows.
\eqn\EffectiveActionii{
\eqalign{
&\ln\Det~M = -D K^2 \sum_{x,\Gm}[
 2\BW^2  -2\Gf^{\Ga}_x \Gf^{\Ga}_{x+\Gm} ] \cr %
&-DK^4 \sum_{x,\Gm}\MO
 \BW^4 (7 d -4) - 6\BW^2 \Gf^{\Ga}_x \Gf^{\Ga}_{x+\Gm} \MC \cr %
&-DK^4 \sum_{x,\Gm}\MO
 2\Gf^{\Ga}_x \Gf^{\Ga}_{x+\Gm}\Gf^{\Gb}_x \Gf^{\Gb}_{x-\Gm}
  - 2\Gf^{\Ga}_{x+\Gm} \Gf^{\Ga}_{x-\Gm}
 + \Gf^{\Ga}_x \Gf^{\Ga}_{x+\Gm}\Gf^{\Gb}_x \Gf^{\Gb}_{x+\Gm}
   - \half   \MC \cr %
&-DK^4 \sum_{\scriptstyle x,\Gm,\Gn
                    \atop \scriptstyle \Gm\neq\Gn}\MO
 4\Gf^{\Ga}_x\Gf^{\Ga}_{x+\Gm}\Gf^{\Gb}_{x+\Gm}\Gf^{\Gb}_{x+\Gm+\Gn}
 - 2\Gf^{\Ga}_x \Gf^{\Ga}_{x+\Gm+\Gn}   \cr %
&
 - \Gf^{\Ga}_x \Gf^{\Ga}_{x+\Gm}\Gf^{\Gb}_{x+\Gn} \Gf^{\Gb}_{x+\Gm+\Gn}
  +\Gf^{\Ga}_x \Gf^{\Ga}_{x+\Gm+\Gn}\Gf^{\Gb}_{x+\Gm}\Gf^{\Gb}_{x+\Gn}
  -\Gf^{\Ga}_x \Gf^{\Ga}_{x+\Gn}\Gf^{\Gb}_{x+\Gm}\Gf^{\Gb}_{x+\Gm+\Gn}
\MC .\cr }}
Thus we have obtained the action of the effective scalar model up to
$O(K^4)$.

  In order to estimate the two- and the four-point correlation functions
of the scalar field, we apply the simplest mean-field (MF) method.
We introduce the external field $H^\Ga~\,\,~(\Ga=0,1,2,3)$
and define the expectation  $\HVEV{\,\, \cdot \,\,}$ by
\eqn\MFT{
\eqalign{
\HVEV{\,\,(\cdot)\,\, }
&= Z_H^{-1}\int\prod_{x,\Ga} d\Gf^\Ga_x
  \exp (\sum_{x,\Ga} H^\Ga \Gf^\Ga_x) \,(\cdot) , \cr %}
%\eqn\MFTii{
 Z_H &= \int\prod_{x,\Ga} d\Gf^\Ga_x
  \exp (\sum_{x,\Ga} H^\Ga \Gf^\Ga_x).\cr}}
In this framework of MF theory, we can compute the correlation function
explicitly, using the fact:
\eqn\MFTiv{
\HVEV{\Gf_x^\Ga}=H^\Ga / 4 +O(H^3),  \,\,\,
\Hvev{\Gf_x^\Ga \Gf_x^\Gb}=\Gd^{\Ga,\Gb} / 4+O(H^2) , }
 \eg,
%\eqn\MFTiii{
$
%\HVEVtrFFFF{0}{\oneh}{\oneh+\twoh}{\oneh}
\AO\GF^{\dag}_0\GF_{\oneh}\GF^{\dag}_{\oneh+\twoh}\GF_{\oneh}\AC
= const. - {H^2 \over 8} + O(H^3) .
$ %}
For the MF free energy given by
\eqn\MFTv{
F(H) = \HVEV{S_{\eff}}
+ \sum_{x,\Ga} H^\Ga \HVEV{\Gf^\Ga_x} - \ln Z_H, }
the stability-violating condition of the PM phase [\EK]
%\eqn\MFTvi{
$
{\pd^2 \over \pd H^2} F(H)|_{H = 0}=0,
$ %}
yields the critical surface separating the FM phase from the PM phase
in the three-dimensional bare parameter space $(\BY,\BW,\BH)$ as
follows.  {\sl FM(S)-PM(S) critical surface}:
\eqn\FmsPms{
\BH_1 ={2\over d}\MKO 1- {\tilde N_f}
 \MO
  K^2 d + K^4 (6d^2 - 3d) \BW^2 + \ha K^4 ( 2 d^2 - d)
 \MC
\MKC ,}
where ${\tilde N_f}=N_f D=N_f 2^{d/2}$. % for convenience
  To obtain the critical surface separating the AFM phase from
the PM phase, on the other hand, we have only to evaluate $\ln\Det~M$ in
terms of the staggered field $\hat \GF$.
Using the translational invariance, we obtain
\eqn\MFTvii{
\eqalign{
&\HVEV{\ln\Det~M} =
-{1\over2}K^2 D V
  d \MO w^2 \tr (1) + \HVEVtrFFh{0}{\oneh} \MC  \cr %
&-{1\over4}K^4 D V
  \MCO 2 d ( 7 d -4 ) \BW^4\tr (1)
  +12 d (2 d - 1 ) \BW^2 \HVEVtrFFh{0}{\oneh}   \cr %
&+8d(d-1) \HVEVtrFFFFh{0}{\oneh}{\oneh+\twoh}{\oneh}
  -4d(d-1) \HVEVtrFFFFh{0}{\oneh}{\oneh+\twoh}{\twoh}   \cr %
&+4d \HVEVtrFFFFh{0}{\oneh}{\oneh+\oneh}{\oneh}
  +2d \HVEVtrFFFFh{0}{\oneh}{0}{\oneh}
 ]  + O(K^6), \cr }}
where $V$ is the volume of the lattice.
Applying the MF theory similarly, we obtain
from the above condition {\sl AFM(S)-PM(S) critical surface}:
\eqn\AfmsPms{
\BH_2 ={2\over d}\MKO -1- {\tilde N_f}
 \MO
  K^2 d + K^4 (6d^2 - 3d) \BW^2 - \ha K^4 ( 2 d^2 - d)
 \MC
\MKC . }

  Now we pay attention to the intersection point of the two  critical
surfaces $\BH=\BH_1$, $\BH=\BH_2$ just obtained.
In the leading order ($O(K^2)$) of HPE, there is no cross point as in the
case of $\BW=0$ %.
 [\EK,
 \ref\ST{ M.A. Stephanov and M.M. Tsypin,
 \PLB{261} (1991) 109.}].
In the next-to-leading order ($O(K^4)$), the cross point does appear
and makes a line in the bare parameter
space $(\BY,\BW,\BH)$, see %Fig.~(a).
 \fig\PDabc{
 Phase diagram of the Yukawa model
 with the Wilson-Yukawa term for $N_f=4$, $d=4$:  \hfil\break %
  (a) schematic three-dimensional phase diagram,  \hfil\break %
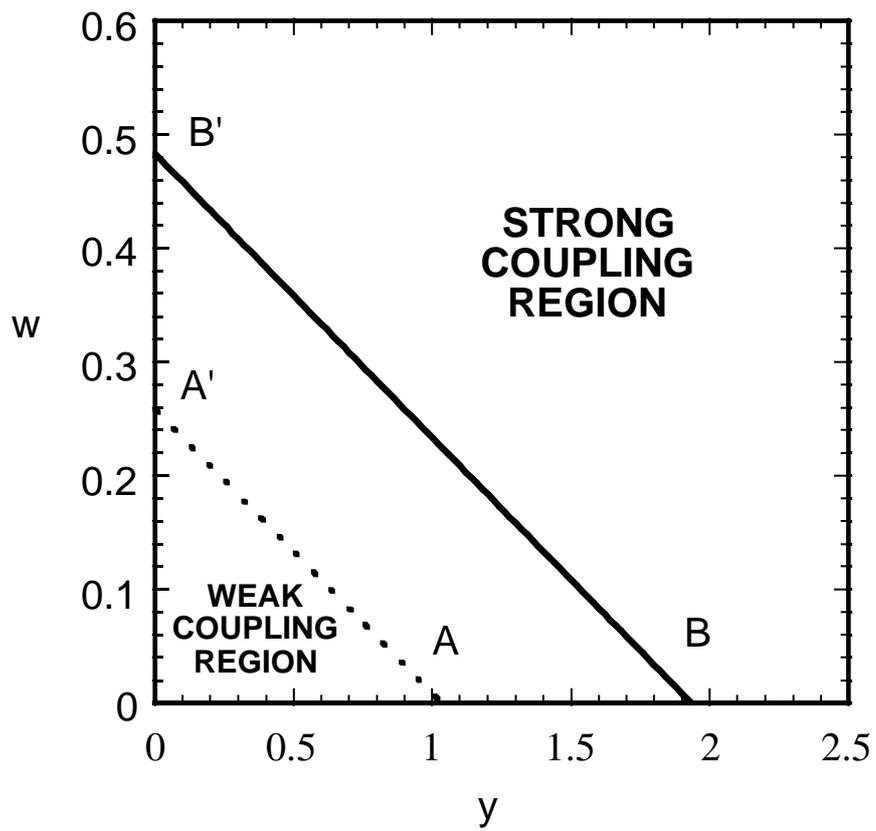
  (b) projection of the phase diagram on
 the $\BH$=constant plane    %,  \hfil\break %
 .}~(a).
This is nothing but the multiple point at which four different
phases meet.
By equating both equation for the critical surface, $\BH_1=\BH_2$, the
intersection line $\BH=\BH_c$ is obtained:
\eqn\CrossB{
\BY+d\BW = {1\over2} \MKO
 {{\tilde N_f}\over 2} (2d^2-d) \MKC^{1/4} ,}
which is depicted in \PDabc~(b).
Indeed, in the limit $\BW\to 0$, this line has the
end-point, $B=(\BY_B,0,\BH_B)$:
%\eqn\CrossBEnd{
$\BY_B = \ha \MKO {{\tilde N_f}\over 2}(2d^2-d)\MKC^{1/4} , \
\BH_B = -{2\over d} \MKO {2{\tilde N_f} d\over 2d-1}\MKC^{1/2}$, % }
which agrees with the previous result [\EK].
Another end-point in the limit $\BY=0$ locates
at $B'=(0,\BW_{B'},\BH_{B'})$:
%\eqn\CrossBEndw{
$
\BW_{B'} = {1\over 2d}\MKO {{\tilde N_f}\over 2}(2d^2-d)\MKC^{1/4} , \
\BH_{B'} = -{2\over d}\MKO {2d{\tilde N_f}\over 2d -1} \MKC^{1/2}
 \MCO 1+{3\over 4d^2}(2d-1) \MCC,
$ %}
where
$\BW_{B'} = 0.48$,         %0.48358415
$\BH_{B'} = -2.84$         %-2.839650695
for $N_f=4$, $d=4$.

%ab
  Although our explicit calculation  in the presence of
the Wilson-Yukawa term ($\BW\neq~0$) is limited up to order $K^4$,  the
existence of the multicritical line is guaranteed up to order $O(K^8)$,
since this line should have in the limit $\BW=~0$ the cross point $B$
whose existence is confirmed up to $O(\BY^{-8})$ in
the previous paper~[\EK].
 Strictly speaking, we can not exclude the case that
the multicritical line may terminate for large $\BW$ and small $\BY$.
However the Monte Carlo simulations~[\Bocki]
show that this is not the case.

 An important aspect of lattice construction which is not mentioned in
this paper concerns the reflection positivity (RP).
So far RP could be proven in a restricted range of the bare parameters in
a restricted class of models
[\ref\LMMW{L. Lin, I. Montvay, G. M\"unster and H. Wittig,
 \NPB{355} (1991) 511.},
\ref\Zenkin{S.V. Zenkin, DFTUZ/92/7.}] .
It is not yet clear whether RP holds also in the model treated in this
paper
[\ref\S{J. Smit, Nucl. Phys. B (Proc. Suppl.) 20 (1991) 542.}]
and in particular in the region which we are focusing on.
\par

%\newsec{Conclusion and discussion}

 In conclusion, we have shown by the analytical method based on
the hopping parameter expansion up to next-to-leading order that
the multicritical line at which four phases meet exist even in the
presence of the Wilson-Yukawa coupling $\BW$ for
the chiral $SU(2)_L\times~SU(2)_R$ symmetric lattice Yukawa model.
This multicritical line is connected to the multicritical point $B$ in
the naive Yukawa model $\BW=0$. In other words, the PM(W) phase and the
PM(S) phase are separated by the intermediate FI phase.
As a result it turns out that the PM(W) phase is not analytically
connected to the PM(S) phase, and the physics in these two respective
phases may be rather different.

  In our model the scalar field is radially fixed which corresponds to
the infinite self-coupling limit $\Gl\rightarrow\infty$ of the scalar
potential, $\Gl (\GF \GF^\dagger-1)^2$.
It is a common wisdom that the radially fixed scalar model belongs to the
same universality class as the weak-coupling  one in the pure scalar sector.
However recent numerical result based on large $N$ shows that at small bare
quartic  self-couplling  $\Gl$  the point $A$ is replaced by a first
order  phase transition line,  before the FM and AFM phases come close
to each other %.
 [\ref\HJS{  A. Hasenfratz, K. Jansen and Y. Shen,
 talk given at Lattice'91.}].
Therefore the multicritical point $B'$ in the limit $\BY\to 0$ is
expected to be only one possible point for
a nontrivial ultraviolet fixed point, since the vanishing of
the fermion mass in lattice units in the limit is guaranteed
from the Golterman-Petcher symmetry
 [\ref\GP{ M.F.L. Golterman and D.N. Petcher,
 \PLB{225} (1989) 159.}]
and the doublers are removed by giving them masses of the order
of the cutoff.

\par
There are however recent works
 [\ref\GPS{ M.F.L. Golterman, D.N. Petcher and J. Smit,
 \NPB{370} (1993) 51;
 D.N. Petcher, talk given at the Conf. Lattice'91, Tsukuba, Japan.},
  \ref\GPR{ M.F.L. Golterman, D.N. Petcher and E. Rivas,
 Wash. U. HEP/91-82.},
 \ref\BDS{ W. Bock, A.K. De and J. Smit, HLRZ-91-81/ITFA-91-30;
W. Bock, talk given at the Conf. Lattice'91, Tsukuba, Japan.}]
which claim that
the continuum limit $\Gk \rightarrow \Gk_c$, $y \rightarrow 0$
from the strong coupling regions FM(S) and PM(S) does most probably not
lead to the desired formulation of the scalar-fermion sector of the
Standard model on the lattice.
  It has been demonstrated in these works that the spectrum in the
PM(S) phase consists only of the scalar particles $(\GF)$ and
a neutral fermion described by the field $\Gy'$.
  Moreover it has been argued that one ends up with an almost non-interacting
neutral fermion in the continuum limit from the FM(S) phase [\GPS],
since the renormalized Yukawa coupling of
the neutral fermion to the scalar particles vanishes as a power of
the lattice spacing $a$.
In this scenario the dimensionless scalar expectation
$\langle \GF_{ren} \rangle$ goes to zero in the continuum limit.
\par
However we do not agree with this scenario and expect another one.
In order to get the theory in the broken-symmetry phase by taking the
scaling continuum limit, bare parameters must be adjusted in such a way
that the vacuum expectation value of the renormalized scalar field
$\langle \GF_{ren} \rangle$
take a non-zero (finite) value in the limit.
To perform this actually, we need detailed information on the phase
structure and the critical behavior of the model as suggested from
experiences in the scalar  $\Gl \Gf^4$ theory.
The scalar  $\Gl \Gf^4$ theory in the broken-symmetry phase constructed
in the above sense has no self-interactions, i.e., the
(dimensionless) renormalized  three- and four-point self-couplings vanish
in the continuum limit
[\ref\Kondo{K.-I. Kondo, Prog. Theor. Phys. 79 (1988) 1217
and references therein.}] .
This is nothing but the triviality of the
broken-symmetry $\Gl \Gf^4$ theory.
Thus we expect a scenario in which the non-vanishing
Yukawa coupling is obtained by keeping the renormalized expectation value
non-zero in the scaling limit by giving the triviality of
the pure scalar sector as a sacrifice.
We do not think that the triviality of the Yukawa coupling has been
established at least in the case of approaching the line B-B' from the
broken-symmetry phase.
Detailed discussion on this issue will be given
in a subsequent paper  %.
 [\ref\EKii{ T. Ebihara and K.-I. Kondo, in preparation.}].

\par
The behavior of the scalar propagator has already been investigated in the
neighborhood of two multicritical points [\BDFJT].
As a result, it  has been shown that the scalar particle spectrum near the
multicritical points A and B  contains beside the usual bosonic
particle additional 'staggered' particles which emerge as a consequence of
the antiferromagnetic ordering at negative $\Gk$ .
The emergence of these spurious particles and the antiferromagnetic phases
has to be simply regarded as an artifact of
the special hypercubic lattice geometry and is a non-universal feature of
this special kind of lattice regularization.
However we do not think that the existence of
two multicritical points A and B may be such a lattice artifact.
This issue will deserve further studies.

\par

\bigbreak\bigskip\centerline{{\bf Acknowledgements}} \nobreak
The authors would like to thank the referee for recalling and informing
them  recent works mentioned in the discussion.

\listrefs
%\vfill\eject

\listfigs
\vfill\eject

\bye